# Microfluidic methods to form artificial cells and to study basic functions of membranes

*Petra S. Dittrich - ETH Zurich*

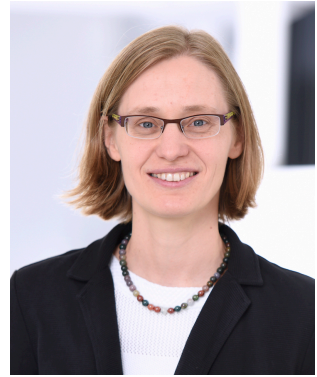

## Biography

*Petra Dittrich is Associate Professor for Bioanalytics at the Department of Biosystems Science and Engineering at ETH Zurich, Switzerland, since 2014. Her research in the field of lab-on-chip-technologies focuses on the miniaturization of high-sensitivity devices for chemical and biological analyses, and microfluidic-aided organization of materials.*

*She studied chemistry at Bielefeld University (Germany) from 1993 to 1999 and Universidad de Salamanca (Spain) in 1997.  After her PhD thesis at the Max Planck-Institute for Biophysical Chemistry (MPI Göttingen, Germany) she stayed for another year as postdoctoral fellow at the MPI Göttingen. Afterwards, she had a postdoctoral appointment at the Institute for Analytical Sciences (ISAS Dortmund, Germany) (2004-2008). In 2008, she became Assistant Professor at the Organic Chemistry Laboratories of the Department of Chemistry and Applied Biosciences (ETH Zurich). Petra Dittrich was fellow of the Studienstiftung des Deutschen Volkes, the German Exchange Organization DAAD and the Christiane Nüsslein-Vollhard-foundation. She was awarded the Analytica Forschungspreis (2010) and the Heinrich-Emmanuel-Merck award (2015). In addition, she received the Starting Grant of the European Research Council (ERC) in 2008 and the ERC Consolidator Grant in 2016.*

## Abstract

Lipid membranes are essential in cellular processes. They define the size of cells and organelles and form a tight barrier. At the same time, the transfer of small molecules and ions into and out of the cell is possible by passive permeation or membrane proteins or processes such as endocytosis or exocytosis. Furthermore, biomechanical forces such as shear stress and weight influence membranes directly, e.g., by deforming their shape. Information about strength and nature of such external forces is transferred into the intracellular side, a process referred to as mechanotransduction. Many of these essential processes have been studied in living cells, however, for a clearer understanding of the underlying mechanisms, model membranes such as lipid vesicles are frequently employed. In the past, we introduced microfluidic methods that enable the creation of vesicles and tubular membrane structures [1] and comprehensive studies with lipid vesicles for the analysis of molecular permeation, fusion and deformation of membranes. The methods are employed to understand fundamental properties of lipid bilayers, and are the scaffolding to form artificial (minimal) cells in a bottom-up approach.

In the recent years, we have developed devices that allowed for immobilization of one or several vesicles and cells in hydrodynamic traps (Fig. 1) [2,3]. These platforms enabled detailed kinetic and mechanistic studies of interactions of molecules with membranes. The combination with high-sensitivity detection techniques based on fluorescence spectroscopy and microscopy proved particularly useful to monitor the effect of peptides on membranes. Peptides are currently intensively discussed as potential drug molecules or drug carriers. They are simple to synthesize yet provide a large structural and functional diversity. We used different types of short peptides that interacted in various ways with the artificial membranes: permeating the membrane; partitioning into the membrane; aggregating inside the membrane; fluidizing or lysing the membrane by formation of pores [4,5]. The reasons for the different properties are under discussion and in the focus of ongoing studies, with the ultimate goal to predict and tailor the properties of the peptides.

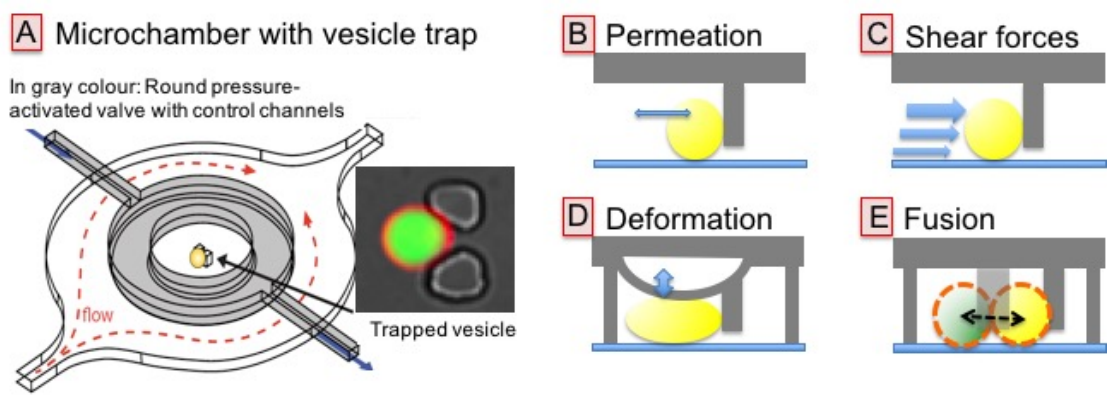

*Figure 1. Studies with lipid vesicles in microfluidic devices. A) Schematics and image of a microchamber with a hydrodynamic trap in the centre. The chamber is closed, when the round valve is activated. The micrograph depicts a captured vesicle, where the membrane and the lumen is stained with fluorescent dyes. B-E) Schematic side view of the trap with a captured vesicle. Various studies were performed to analyse permeation or partitioning of molecules, the effect of shear forces and deformation on the lipid membrane as well as the mechanism of membrane fusion.*

In addition, we used the microfluidic device to analyse the effect of shear stress on membranes [6]. For this, we created vesicles that form two types of lipid domains that can be stained with different dyes, and exposed them to fluid flow. In this way, the transient, non-deterministic patterns of lipid domains could be visualized. After stopping the flow, the lipid domains immediately return to round-shaped patches and fuse to larger domains. However, not only the membrane is effected by the fluid flow. We also visualized the flow inside the vesicle by means of fluorescent tracer particles. The movements of these particles were monitored by defocussing microscopy and analysed with a self-written algorithm. The results reveal the two-hemispheric flow profile inside the vesicles. Interestingly, any asymmetries of the external shear forces transfer into asymmetric bi-hemispheric patterns in the lumen. Together, these findings also indicate a higher fluidity of the membrane under flow conditions and a strong transduction efficiency across the membrane.

Small variations of the microfluidic device were made to deform lipid vesicles and image the changes of the membrane in response to these mechanical strains. For these measurements, an additional stamp was implemented that could be deflected on top of a vesicle, while the membrane is monitored by confocal laser scanning microscopy. In another modification, we changed the design of the hydrodynamic trap to capture two vesicles. The tight contact between the membrane allowed for efficient fusion of the vesicle pairs, induced either by an electrical pulse [7] or by means of fusogenic peptides. Moreover, fusion studies of liposomes and natural cells were conducted [8].

Finally, we employ microfluidic methods for mimicking artificial cells [9,10]. With bulk methods, it is challenging to reproduce cellular architecture, where smaller lipid compartments are enclosed within the cytosol.

In summary, our microfluidic devices facilitate comprehensive studies on membranes with high resolution imaging techniques. Both the technological improvements of the platform as well as the acquired knowledge about membranes will open the way to produce and tailor membranes on demand and hence, to mimic living cells in future.

## Acknowledgement
Financial support of the European Research Council ("HybCell", ERC Consolidator Grant no. 681587) is gratefully acknowledged.